\documentclass[aps, prb, superscriptaddress, reprint, nobibnotes, twocolumn]{revtex4-2}
\usepackage{amsmath,amssymb}
\usepackage{color}
\usepackage{comment}
\usepackage{bm,graphicx,hyperref}
\hypersetup{%
  breaklinks = {true},
  citecolor = {blue},
  colorlinks = {true},
  linkcolor = {red},}

\usepackage{cancel}

\begin{document}

\title{Time-local stochastic equation of motion for solid ionic electrolytes}

\author{A. Rodin}
\affiliation{Yale-NUS College, 16 College Avenue West, 138527, Singapore}
\affiliation{Department of Materials Science and Engineering, National University of Singapore, 117575, Singapore}

\author{B. A. Olsen}
\affiliation{Department of Physics, Lewis \& Clark College, Portland, Oregon, 97219, USA}

\author{A. Ustyuzhanin}
\affiliation{Constructor University, Bremen, Campus Ring 1, 28759, Germany}
\affiliation{Institute for Functional Intelligent Materials, National University of Singapore, 4 Science Drive 2, Singapore 117544, Singapore}

\author{A. Maevskiy}
\affiliation{Institute for Functional Intelligent Materials, National University of Singapore, 4 Science Drive 2, Singapore 117544, Singapore}

\begin{abstract}

Numerical studies of ionic motion through solid electrolytes commonly involve static nudged-elastic band (NEB) methods or costly \emph{ab initio} molecular dynamics (AIMD).
Building on a time-local model of current carrier-electrolyte interaction and incorporating thermal motion, we introduce an approach that is intermediate between the two well-established methodologies by treating the electrolyte as an effective medium that interacts with the mobile particle.
Through this coupling, the thermally vibrating electrolyte imparts energy to the charge carriers while also absorbing energy from them due to its own finite elasticity.
Using a simple model system, we validate our approach through a series of numerical simulations.
Our methodology reproduces both dissipative and diffusive behavior, and helps link microscopic system parameters to measurable macroscopic properties.

\end{abstract}	

\maketitle

\section{Introduction}
\label{sec:Introduction}

Ionic transport through solids~\cite{MahanRoth, Mehrer} is a fundamental physics problem at the root of solid-state batteries~\cite{Bachman2016, Manthiram2017, Famprikis2019}, hydrogen fuel cells~\citep{Talukdar2024}, electrolysis cells~\citep{Wolf2023}, and electrochemical synapses~\citep{Huang2023}.
This transport consists of current-carrying mobile ions traveling through an electrolyte whose constituent atoms remain close to their equilibrium positions, retaining the structural integrity of the material.
Compared to their liquid counterparts, solid electrolytes can enhance device robustness by suppressing dendrite formation~\citep{Famprikis2019} and operate over a larger temperature range~\citep{Bachman2016, Manthiram2017}.
Microscopically, the mobile conducting ions intermittently become trapped in, and escape from, local potential energy minima within the electrolyte, leading to macroscopic limits on conductivity.
Despite solid-state electrolytes' advantages, their conductivity is currently smaller then traditional liquid electrolyte technology~\citep{Bachman2016}, so substantial effort is being dedicated to finding materials with better transport properties.
One avenue in this effort is to find materials that minimize the amount of ``trapped'' time the mobile ions spend in potential wells.

In recent years, the search for better electrolyte materials has been dominated by various \emph{ab initio} methods, which provide a glimpse into the microscopic processes of ionic conduction.
While some efforts have focused on which material properties facilitate escape from potential minima~\citep{Wang2015, Bo2016, Krauskopf2017, Brenner2020}, others have explored conduction mechanisms that reduce the chance of ions becoming trapped~\citep{Deng2015, Marcolongo2017, He2017, DiStefano2019, Fall2019, Poletayev2022, Carvalho2022, Hu2023}.
The most common computational approaches---nudged elastic band (NEB) calculations and \emph{ab initio} molecular dynamics (AIMD) simulations---have somewhat complementary strengths and shortcomings.

In NEB calculations, the mobile ion location varies from one energy minimum to another in a series of steps---at each location, the ion is held fixed along the transport direction while the rest of the system relaxes.
The total energy is calculated along the ion's path, and the maximum is the activation energy $E_a$.
This quasistatic NEB approach to estimate $E_a$ can help efficiently suggest material classification.
A common technique uses $E_a$ to predict charge carrier number (and, by extension, conductivity) scaling with temperature using the Arrhenius form: $e^{-E_a / k_BT}$.
However, recent work~\citep{Noori2024} has shown that this Arrhenius dependence can break down even when the ``electrolyte" consists of a single one-dimensional oscillator.
Moreover, since NEB calculations are quasistatic, they do not capture the dynamical effects that determine the prefactor for the Arrhenius term.

AIMD computes the trajectories of each of the framework atoms and the mobile ions in the presence of an externally-induced thermal motion.
These simulations make it possible to estimate mobility either using the Nernst-Einstein relation and numerically-determined diffusivity, or directly using an applied bias voltage~\citep{Carvalho2022}.
While the evolution in these trajectories is governed by realistic forces calculated from first principles with density functional theory (DFT), this approach is very computationally costly.
To keep computational resources manageable, a typical calculation evolves the system at extremely high temperatures for only several nanoseconds, and is limited to a few unit cells.
For similar reasons, direct computations of the mobility require extreme electric field strengths (e.g., Ref.~\citep{Carvalho2022} used $0.075\text{~V/\AA}=7.5\times 10^8$~V/m, about 250 times higher than the breakdown voltage of air). 

In this paper, we use microscopic theory to construct a stochastic single-particle equation to describe the motion of a mobile ion.
This formalism incorporates dynamical effects, which do not arise in NEB calculations, and can evolve under more realistic system parameters than are feasible with AIMD computations.
At the cost of using less realistic potentials, this tool can complement the other methods as part of a multi-pronged approach to identify promising electrolyte candidates.
Its primary utility comes from allowing dynamical simulations at longer timescales, and over a greater temperature range, leading to improved statistics and direct prediction of temperature scaling.

This paper is organized as follows: we give a microscopic description of ionic motion through a solid electrolyte in Sec.~\ref{sec:Model}.
There, we show that the displacement of the electrolyte's atoms in response to the mobile ions is given by a transcendental time-local Eq.~\eqref{eqn:r_by_parts_local}.
Next, we introduce a model system in Sec.~\ref{sec:Model_system} and validate our time-local treatment in Sec.~\ref{sec:Time_locality_validation}, where we show that additional simplifications remove the transcendental nature of Eq.~\eqref{eqn:r_by_parts_local}.
We demonstrate our formalism's power to capture dissipation and diffusion in Secs.~\ref{sec:Dissipation} and \ref{sec:Diffusion}, respectively.
In section~\ref{sec:Discussion}, we discuss of the results, assumptions, limitations, and consequences of our study, and conclude with a summary in Sec.~\ref{sec:Summary}.

\section{Model}
\label{sec:Model}

As in our earlier work,~\citep{Rodin2022, Rodin2022a, Mahalingam2023, Mahalingam2023a, Rodin2024b} we start with a Lagrangian describing the motion of a particle of mass $M$ through a crystalline framework:

\begin{align}
    L &= 
    \frac{M}{2}\dot{\mathbf{R}}^T\dot{\mathbf{R}} 
    +  \frac{1}{2}\dot{\mathbf{r}}^T\tensor{m}\dot{\mathbf{r}}
    -  \mathbf{r}^T \frac{\tensor{V}}{2}\mathbf{r}
    - U(\mathbf{r},\mathbf{R})\,.
    \label{eqn:Lagrangian}
\end{align}
Here, $\mathbf{R}$ is the mobile particle's position, $\mathbf{r}$ is a vector of framework atom displacements from their equilibria, and $\tensor{m}$ is the framework's mass matrix.
$\mathbf{r}^T \tensor{V} \mathbf{r} / 2$ gives the framework's potential energy in the harmonic approximation and $U(\mathbf{r}, \mathbf{R})$ is the interaction between the framework and particle.

Equation~\eqref{eqn:Lagrangian} yields the standard equations of motion

\begin{align}
    M\ddot{\mathbf{R}} &= - \nabla_{\mathbf{R}} U\left(\mathbf{r},\mathbf{R}\right)\,,
    \label{eqn:Mobile_EOM}
    \\
   \tensor{m}\ddot{\mathbf{r}} &= -\tensor{V}\mathbf{r} -\nabla_{\mathbf{r}} U\left(\mathbf{r},\mathbf{R}\right)\,.
    \label{eqn:Framework_EOM}
\end{align}
Although Eqs.~\eqref{eqn:Mobile_EOM} and \eqref{eqn:Framework_EOM} contain all the information necessary to solve the problem, they are unmanageable for large systems.
Fortunately, it is possible to write down a formal solution for $\mathbf{r}$ which can then be used to solve Eq.~\eqref{eqn:Mobile_EOM}.
We start by writing Eq.~\eqref{eqn:Framework_EOM} as a symmetric eigenvalue problem 

\begin{equation}
     \tensor{m}^\frac{1}{2}\ddot{\mathbf{r}} =-0^+\dot{\mathbf{r}} -\tensor{m}^{-\frac{1}{2}}\tensor{V} \tensor{m}^{-\frac{1}{2}} 
     \left(\tensor{m}^\frac{1}{2}\mathbf{r}\right)
     -
     \tensor{m}^{-\frac{1}{2}}\nabla_{\mathbf{r}} U\left(\mathbf{r},\mathbf{R}\right)\,,
    \label{eqn:Framework_EOM_Symmetric}
\end{equation}
where $0^+$ is an infinitesimal dissipation.
Without the final term, $\mathbf{r}$ describes a homogeneous solution for the framework which corresponds to thermal vibration.
We will address this motion component below and focus on the framework's response to its interaction with the particle first.
Taking the Fourier transform of Eq.~\eqref{eqn:Framework_EOM_Symmetric} with respect to time and solving for $\mathbf{r}_\omega$ gives

\begin{align}
    \mathbf{r}_\omega &=
    \tensor{m}^{-\frac{1}{2}}
    \left(\omega^2+i\omega 0^+ -\tensor{m}^{-\frac{1}{2}}\tensor{V} \tensor{m}^{-\frac{1}{2}}\right)^{-1}
    \tensor{m}^{-\frac{1}{2}}
    \nonumber
    \\
    &\times
   \mathcal{F}\left[\nabla_{\mathbf{r}} U\left(\mathbf{r},\mathbf{R}\right)\right]\,,
    \label{eqn:Framework_EOM_Symmetric_FT}
\end{align}
where $\mathcal{F}[\dots]$ denotes the Fourier transform.
Because $\tensor{m}^{-\frac{1}{2}} \tensor{V} \tensor{m}^{-\frac{1}{2}}$ is a real symmetric matrix, there exists a matrix $D = \left[\boldsymbol{\varepsilon}_1, \boldsymbol{\varepsilon}_2, \dots\right]$ such that $D^\dagger\tensor{m}^{-\frac{1}{2}} \tensor{V} \tensor{m}^{-\frac{1}{2}}D = \tensor{\Omega}^2$, where $\tensor{\Omega}^2$ is a diagonal matrix and $\boldsymbol{\varepsilon}_j$ are eigenstates of $\tensor{m}^{-\frac{1}{2}} \tensor{V}  \tensor{m}^{-\frac{1}{2}}$.
Thermally-excited eigenmodes produce the homogeneous framework motion referred to above. 
Taking the eigenvalues for the $j$th state to be $\Omega_j^2$, we have

\begin{equation}
    \left(\omega^2+i\omega 0^+-\tensor{m}^{-\frac{1}{2}} \tensor{V} \tensor{m}^{-\frac{1}{2}}\right)^{-1}
    =
    \sum_j \frac{\boldsymbol{\varepsilon}_j\boldsymbol{\varepsilon}_j^\dagger}{(\omega +i0^+)^2- \Omega_j^2}\,,
\end{equation}
leading to

\begin{align}
    \mathbf{r}&= 
    \sum_j\mathcal{F}^{-1}\left[ \frac{    \tensor{m}^{-\frac{1}{2}}\boldsymbol{\varepsilon}_j\boldsymbol{\varepsilon}_j^\dagger    \tensor{m}^{-\frac{1}{2}}}{(\omega +i0^+)^2 - \Omega_j^2}
    \right]
    \ast
     \frac{ \nabla_{\mathbf{r}} U\left(\mathbf{r},\mathbf{R}\right)}{\sqrt{2\pi}}
  \,,
    \label{eqn:Framework_EOM_Symmetric_FT_Solution}
\end{align}
where $\ast$ denotes the convolution.
Using

\begin{align}
   \frac{1}{\sqrt{2\pi}}\mathcal{F}^{-1}\left[ \frac{1}{(\omega +i0^+)^2- \Omega_j^2}
    \right]
    =&
   - \frac{\sin\Omega_j t }{\Omega_j}
    \Theta(t)\,,
\end{align}
where $\Theta(t)$ is the Heaviside step function, we have

\begin{align}
    \mathbf{r}(t)&=
    -\int_{0}^tdt' \frac{d\tensor{W}(t-t')}{dt'}
   \nabla_{\mathbf{r}} U\left[\mathbf{r}(t'),\mathbf{R}(t')\right]\,,
    \label{eqn:Framework_Solution}
    \\
    \tensor{W}(t) &= \tensor{m}^{-\frac{1}{2}}
     \sum_j \boldsymbol{\varepsilon}_j \boldsymbol{\varepsilon}_j^\dagger \frac{\cos\left(\Omega_jt\right)}{\Omega_j^2}
   \tensor{m}^{-\frac{1}{2}}\,.
   \label{eqn:W}
\end{align}
Having obtained the particular solution determined by the forcing term, we add the homogeneous trajectory $\mathbf{r}_H(t)$ to Eq.~\eqref{eqn:Framework_Solution}.
Integrating the forcing term by parts gives
\begin{align}
       \mathbf{r}(t)
   &=  \mathbf{r}_H(t)
   -
    \overbrace{ \tensor{W}(0)}^{\tensor{V}^{-1}}
   \nabla_\mathbf{r}U[\mathbf{r}(t),\mathbf{R}(t)]
   \nonumber
   \\
   &+
   \tensor{W}(t)
   \nabla_\mathbf{r}U[\mathbf{r}(0),\mathbf{R}(0)]
   \nonumber
   \\
   &+
   \int^t_0 dt'
  \tensor{W}(t-t')
   \frac{d}{dt'}\nabla_\mathbf{r}U[\mathbf{r}(t'),\mathbf{R}(t')]
   \,.
   \label{eqn:r_by_parts}
\end{align}
The third term on the right-hand side goes to zero at large $t$ because it is a sum of oscillating cosine functions.
Therefore, we will drop it in the subsequent analysis.
If the time derivative of the force vanishes, we can introduce $\boldsymbol{\delta}\equiv \mathbf{r} - \mathbf{r}_H$ so that Eq.~\eqref{eqn:r_by_parts} becomes $\boldsymbol{\delta} = -\tensor{V}^{-1} \nabla_\mathbf{r}U[\mathbf{r}_H+\boldsymbol{\delta},\mathbf{R}]$.
This expression corresponds to a fully relaxed static framework configuration for a particular $\mathbf{R}$ with the equilibrium framework positions shifted by $\mathbf{r}_H$, reminiscent of the quasistatic NEB formulation.
Therefore, we can view the final term in Eq.~\eqref{eqn:r_by_parts} as the dynamic correction to the static NEB result.

At this point, we have a formal solution for the framework motion, having avoided solving Eq.~\eqref{eqn:Framework_EOM} for an infinitely large number of degrees of freedom.
In fact, because our main goal is the particle trajectory, we can  focus only on framework atoms that are sufficiently close to the particle to exhibit a non-negligible force, making the size of $\mathbf{r}$ finite.
Unfortunately, even with this simplification, the expression for $\mathbf{r}$ involves a memory integral.
Therefore, our next goal is to convert this expression into an approximate time-local form.

We start by noting that, for a system containing $A$ atoms per unit cell, the eigenvectors $\boldsymbol{\varepsilon}_j$ are

\begin{align}
    \boldsymbol{\varepsilon}_j = 
    \frac{1}{\sqrt{N}}
    \begin{pmatrix}
        e^{i\mathbf{L}_1\cdot\mathbf{q}_j}
        \\
         e^{i\mathbf{L}_2\cdot\mathbf{q}_j}
        \\
        \vdots
        \\
        e^{i\mathbf{L}_N\cdot\mathbf{q}_j}
    \end{pmatrix}\otimes \boldsymbol{\eta}_j
   \,,
    \label{eqn:epsilon}
\end{align}
where $N$ is the number of unit cells in the system, $\mathbf{q}_j$ is the crystal momentum corresponding to the mode $j$, $\mathbf{L}_n$ is the coordinate of the $n$th unit cell, and $\boldsymbol{\eta}_j$ is a $3A$-dimensional eigenvector of the dynamical matrix~\citep{AshcroftMermin}.

To proceed, we assert that low-frequency, long-wavelength (small $\mathbf{q}_j$) modes play the dominant role in the response kernel in Eq.~\eqref{eqn:W}.
This low-energy phonon dominance is due to suppressed coupling to high-frequency modes: the $\Omega_j^2$ in the denominator.
Moreover, even though the number of modes grows with frequency due to the increasing phase space, their short wavelength makes them much more susceptible to destructive interference.
This assertion yields two consequences.
First, we neglect the phase difference between all the unit cells close to the mobile ion and replace $e^{i\mathbf{L}_n\cdot\mathbf{q}_j}\rightarrow 1$ in $\boldsymbol{\varepsilon}_j$.
Second, the true form of the high-energy dispersion is unimportant, allowing us to linearize the spectrum.

To linearize the spectrum, we start with the dynamical matrix 

\begin{align}
     \mathcal{D}(\mathbf{q})
    &=
   \frac{1}{N} \sum_{ab}\left[\tensor{m}^{-\frac{1}{2}} \tensor{V} \tensor{m}^{-\frac{1}{2}}\right]_{ab}e^{i(\mathbf{L}_a -\mathbf{L}_b)\cdot \mathbf{q}
    }
    \nonumber
    \\
    &\equiv
    \sum_{\mathbf{L}}\left[\tensor{m}^{-\frac{1}{2}} \tensor{V} \tensor{m}^{-\frac{1}{2}}\right]_{\mathbf{L}}e^{i\mathbf{L}\cdot \mathbf{q}
    }\,
    \label{eqn:Dynamical_Matrix}
\end{align}
where $\left[\tensor{m}^{-\frac{1}{2}} \tensor{V} \tensor{m}^{-\frac{1}{2}}\right]_{ab}$ are $3A\times 3A$ blocks coupling unit cells at $\mathbf{L}_a$ and $\mathbf{L}_b$.
The last equivalence holds because of lattice periodicity so that $\left[\tensor{m}^{-\frac{1}{2}} \tensor{V} \tensor{m}^{-\frac{1}{2}}\right]_{ab}$ depends on the difference $\mathbf{L}_a - \mathbf{L}_b$ with $\left[\tensor{m}^{-\frac{1}{2}} \tensor{V} \tensor{m}^{-\frac{1}{2}}\right]_\mathbf{L}$ corresponding to the coupling block for two unit cells separated by vector $\mathbf{L}$.
Next, we write $\mathcal{D}(\mathbf{q}) \rightarrow \mathcal{D}_0(\theta, \phi) + q^2\mathcal{D}_1 (\theta, \phi)$, where $\theta$ and $\phi$ are the polar and azimuthal angles of $\mathbf{q}$, respectively.
The three zero-frequency eigenstates of $\mathcal{D}_0(\theta,\phi)$ are then labeled by the angles and the acoustic branch index $u$: $\boldsymbol{\eta}_{u,\theta,\phi}$.
Because the perturbation term $q^2\mathcal{D}_1 (\theta, \phi)$ is second-order in $q$, the eigenstates $\boldsymbol{\eta}_{u,\theta,\phi}$ are unchanged up to the first order in $q$.
The frequency, on the other hand, is $q\sqrt{\boldsymbol{\eta}_{u,\theta,\phi}^\dagger \mathcal{D}_1(\theta,\phi)\boldsymbol{\eta}_{u,\theta,\phi}}\equiv q v_{u,\theta,\phi}$, where $v_{u,\theta,\phi}$ is the direction-dependent speed of sound of the $u$th phonon branch.
If $\mathcal{D}_0(\theta,\phi)$ vanishes, $\boldsymbol{\eta}_{u, \theta,\phi}$ are the eigenstates of $\mathcal{D}_1(\theta,\phi)$ and $q\sqrt{\boldsymbol{\eta}_{u,\theta,\phi}^\dagger \mathcal{D}_1(\theta,\phi)\boldsymbol{\eta}_{u,\theta,\phi}}\equiv q v_{u,\theta,\phi}$ still holds.

In the $q \rightarrow 0$ limit, all atoms in in unit cell move in the same direction with the same amplitude so that

\begin{align}
    \tensor{m}^{-\frac{1}{2}} \boldsymbol{\varepsilon}_j &\approx
      \frac{1}{\sqrt{m N}}
  \underbrace{\begin{pmatrix}
    \tensor{1}_3 \\ \tensor{1}_3 \\ \vdots
\end{pmatrix}}_{\mathcal{K}} \boldsymbol{\psi}_{u,\theta,\phi}
\,,
\label{eqn:simplified_epsilon}
\end{align}
where $m$ is the total mass of the unit cell, $\mathcal{K}$ is a column of $NA$ copies of the $3\times 3$ identity matrix  $\tensor{1}_3$, and $\boldsymbol{\psi}_{u,\theta,\phi}$ is the three-dimensional phonon polarization vector.
Using these simplifications, we write Eq.~\eqref{eqn:W} as

\begin{align}
    \tensor{W}(t) 
  & \simeq 
   \mathcal{K}
          \frac{\mathcal{V}/m}{8\pi^3}\sum_u \int d\mathbf{q}
          \boldsymbol{\psi}_{u,\theta,\phi} \boldsymbol{\psi}_{u,\theta, \phi}^\dagger \frac{\cos\left(q v_{u,\theta,\phi}t\right)}{q^2v_{u,\theta,\phi}^2}
    \mathcal{K}^T
    \nonumber
    \\
   &=
    \mathcal{K}
    \sum_u \int_0^Q 
\frac{dq}{8\pi^3\rho} \int d\mathcal{S}
      \boldsymbol{\psi}_{u,\theta,\phi} \boldsymbol{\psi}_{u,\theta, \phi}^\dagger  \frac{\cos\left(q v_{u,\theta,\phi}t\right)}{v_{u,\theta,\phi}^2}
    \mathcal{K}^T
       \nonumber
\\
   &\simeq 
    \mathcal{K}
         \underbrace{ \sum_u \int 
          \frac{d\mathcal{S}}{8\pi^2\rho}
          \boldsymbol{\psi}_{u,\theta,\phi} \boldsymbol{\psi}_{u,\theta, \phi}^\dagger \frac{\delta\left(t\right)}{v_{u,\theta,\phi}^3}}_{2\mathcal{L}\delta(t)}
   \mathcal{K}^T
          \,,
          \label{eqn:local_kernel}
\end{align}
where $\mathcal{V}$ is the unit cell volume, $\rho = m / \mathcal{V}$ is the material's density, and $d\mathcal{S}$ denotes the integral over the solid angle.
The maximum momentum $Q$ is similar to the cutoff appearing in the Debye model for a linearized spectrum.
In the last line, we took the limit $Q\rightarrow\infty$ to get the time-local expression.
Physically, this approximation means that the relevant dynamics of the system are substantially slower than the fastest mode.
The $3\times 3$ matrix $\mathcal{L}$ gives the approximate time-local response in the linear-spectrum approximation.  

Inserting Eq.~\eqref{eqn:local_kernel} into Eq.~\eqref{eqn:r_by_parts} and taking the time integral gives a time-local transcendental equation for $\mathbf{r}$

\begin{align}
       \mathbf{r}
   &\approx  \mathbf{r}_H
   -
    \tensor{V}^{-1}
   \nabla_\mathbf{r}U(\mathbf{r}, \mathbf{R})
    +
   \mathcal{K}
   \mathcal{L}
  \mathcal{K}^T
   \nabla_\mathbf{r}\frac{d}{dt}U(\mathbf{r}, \mathbf{R})
   \nonumber
   \\
    &=  \mathbf{r}_H
   -
    \tensor{V}^{-1}
   \nabla_\mathbf{r}U(\mathbf{r}, \mathbf{R})
    -
   \mathcal{K}
   \mathcal{L}
   \nabla_\mathbf{R}\frac{d}{dt}U(\mathbf{r}, \mathbf{R})
   \,,
   \label{eqn:r_by_parts_local}
\end{align}
where we used the fact that $\mathcal{K}^T \nabla_\mathbf{r} f(\mathbf{r}, \mathbf{R}) =\nabla_\mathbf{r}f(\mathbf{r}, \mathbf{R}) \mathcal{K} =\sum_{jk}\nabla_{\mathrm{r}_{j,k}}f(\mathbf{r}, \mathbf{R}) = - \nabla_\mathbf{R}f(\mathbf{r}, \mathbf{R})$ to replace the gradient with respect to $\mathbf{r}$ applied on $U$.
In other words, the uniform shift of the framework generated by $\mathcal{K}$ is physically equivalent to a shift of the mobile particle in the opposite direction.
Our aim is to use the time-local form in Eq.~\eqref{eqn:Mobile_EOM} to obtain the mobile particle's trajectory.
First, however, we validate the time-local approximation and build intuition by performing a set of numerical experiments using a model system introduced in Sec.~\ref{sec:Model_system}.
We then show, in Sec.~\ref{sec:Time_locality_validation}, that we can simplify Eq.~\eqref{eqn:r_by_parts_local} by replacing $\mathbf{r}\rightarrow\mathbf{r}_H$ on the right-hand side, leading to

\begin{equation}
    \mathbf{r} = \mathbf{r}_H
   -
    \tensor{V}^{-1}
   \nabla_\mathbf{r}U(\mathbf{r}_H, \mathbf{R})
    -
   \mathcal{K}
   \mathcal{L}
   \nabla_\mathbf{R}\frac{dU(\mathbf{r}_H, \mathbf{R})}{dt}\,,
   \label{eqn:r_simple}
\end{equation}
allowing us to write down the differential equation for $\mathbf{R}$

\begin{align}
    M\ddot{\mathbf{R}} &= -\nabla_\mathbf{R}U(\mathbf{r}_\mathrm{eff} ,\mathbf{R}_\mathrm{eff} )\,,
    \label{eqn:EOM_Simple}
   \\
   \mathbf{r}_\mathrm{eff} & =  \mathbf{r}_H
   -
    \tensor{V}^{-1}
   \nabla_\mathbf{r}U(\mathbf{r}_H,\mathbf{R})\,,
   \label{eqn:r_eff}
   \\
   \mathbf{R}_\mathrm{eff} &=\mathbf{R}+
   \mathcal{L}
   \nabla_\mathbf{R}\frac{dU(\mathbf{r}_H, \mathbf{R})}{dt}\,,
   \label{eqn:R_eff}
\end{align}
where the last term of Eq.~\eqref{eqn:r_simple} was moved to $\mathbf{R}$ inside $U$ because it corresponds to a uniform framework shift.

\section{Model system}
\label{sec:Model_system}

The aim of this section is to introduce a tractable model system and justify our parameter choices.

Typical scales for physical quantities in solid materials are meV for kinetic energies, \AA\ for lengths, and ps for times.
Normalizing our system parameters by these quantities implies a mass scale, where $[m]=[E]/[v^2]=[E][t]^2/[\ell]^2$. 
A dimensionless mass $M=1$ then corresponds to $1$~meV ps$^2/$\AA$^2\approx9.66$~Da.
One of the most common mobile ion species, Li, then has mass $M\approx 0.7$, and the lattice (most commonly formed from Si, Ge, S, and P) then has mass $m\approx 3.5$.

\begin{figure}
    \centering
    \includegraphics[width = \columnwidth]{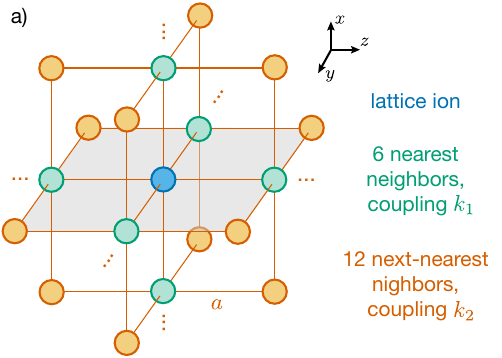}
    \includegraphics[width = \columnwidth]{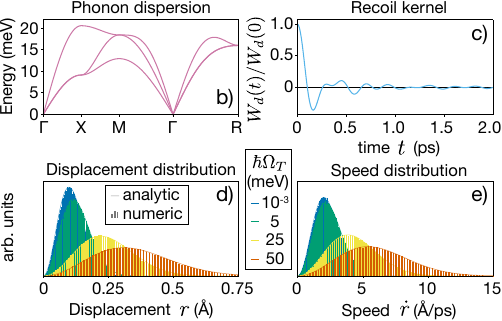}
    \caption{\emph{Model system.} A simple cubic lattice, depicted in (a), where each lattice ion couples to its 6 nearest neighbors (shown in green) with a coupling $k_1$, and to its 12 next-nearest neighbors (shown in orange) with a coupling $k_2$. (b) The phonon dispersion computed using the dynamical matrix in Eq.~\eqref{eqn:D} for $m = 3.5$~meV\,ps$^2/$\AA$^2$, $a = 3$~\AA, $k_1 = 520$~meV/\AA$^2$, and $k_2 =170$~meV/\AA$^2$. (c) Diagonal element of the recoil matrix $\tensor{W}(t)$ for this model system. The rapid decay of this term provides support for the time-local treatment. (d),(e) Distributions of displacements and speeds of the framework ions due to thermal motion at different temperatures. Solid curves are analytic Maxwell-Boltzmann distributions and shaded areas are histograms obtained from numerically generated displacements for a system of $60\times60\times 60$ unit cells.}
    \label{fig:Framework}
\end{figure}

The simplest three-dimensional crystal is a cubic lattice containing a single atom per unit cell, shown in Fig.~\ref{fig:Framework}(a).
A typical nearest-neighbor distance for monoelemental simple cubic, bcc, and fcc lattices is about $3$~\AA, or $a\approx 3$.
This size is within the range of previous benchmarking studies of sulfur model solid electrolytes, with volume per atom of 20--70~\AA$^3$~\citep{Wang2015}.
We stress that cubic lattices generally do not make good ionic conductors, but we chose this simple geometry for demonstration purposes only.

Each atom is coupled to its nearest (next-nearest) neighbors via springs with force constants $k_1$ ($k_2$) so that the dynamical matrix is

\begin{widetext}
\begin{align}
    \mathcal{D}(\mathbf{q}) &= 
    2\frac{k_1}{m}\begin{pmatrix}
        1 - \cos q_x a&0&0
        \\
        0&1 - \cos q_y a&0
        \\
       0& 0& 1 - \cos q_z a
    \end{pmatrix}
    \nonumber
    \\
    &+2\frac{k_2}{m}\begin{pmatrix}
      2 - \cos q_x a \cos q_y a- \cos q_x a \cos q_z a&\sin q_x a \sin q_y a & \sin q_x a \sin q_z a
        \\
         \sin q_y a \sin q_x a& 2 - \cos q_y a \cos q_x a- \cos q_y a \cos q_z a  &  \sin q_y a \sin q_z a
        \\
        \sin q_z a\sin q_x a&  \sin q_z a \sin q_y a&  2 - \cos q_z a \cos q_x a- \cos q_z a \cos q_y a
    \end{pmatrix}\,,
    \label{eqn:D}
\end{align}
\end{widetext}
with $q_{x,y,z} = 2\pi \left[1,2,\dots,N_{x,y,z}\right] / N_{x,y,z}a$.
Keeping $k_2\neq 0$ is important because, in its absence, the Cartesian coordinates become decoupled, decomposing the lattice into individual one-dimensional systems.

The highest phonon angular frequency in the system considered here is $\Omega_\mathrm{max} = \sqrt{4k_1 / m + 8k_2 / m}$, corresponding to the X point in the Brillouin zone.
Taking a reasonable maximum cycle frequency $f_\mathrm{max}= 5\mathrm{~THz}$, corresponding to $\approx 20\mathrm{~meV}$, we have $\Omega_\mathrm{max} = 2\pi f_\mathrm{max} = 10\pi\mathrm{~ps}^{-1}$.
The choice of 5~THz is on the lower end of frequencies found in softer monoelemental systems or two-element ionic conductors (such as lithium halides~\citep{Wang2020a}), picked to get a physically appropriate speed of sound.
For a reasonable value $k_1 / k_2 \approx 3$, we then have $k_2 = 5\pi^2 m \approx 170$~meV/\AA$^2$ and $k_1 \approx 520$~meV/\AA$^2$.
These values translate to $\approx 15\pi$~\AA/ps ($\approx 4700$~m/s) as the speed of sound for the longitudinal phonon along the lattice's principal axes, a physically reasonable value.
The phonon dispersion obtained by diagonalizing Eq.~\eqref{eqn:D} is given in Fig.~\ref{fig:Framework}(b).

An important component of our derivation in Sec.~\ref{sec:Model} is the rapid decay of the recoil kernel $\tensor{W}$ in Eq.~\eqref{eqn:W}.
For the monoatomic lattice considered here, the $3\times 3$ component of this kernel coupling framework atoms $j$ and $k$ is

\begin{align}
   \tensor{W}_{j,k}(t) &= 
     \sum_l  \frac{\boldsymbol{\eta}_l \boldsymbol{\eta}_l^\dagger}{m\Omega_l^2}\frac{e^{i\mathbf{q}_l\cdot (\mathbf{L}_j - \mathbf{L}_k)}}{N}\cos (\Omega_l t)\,.
     \label{eqn:W_cubic}
\end{align}
To illustrate this decay, we plotted its diagonal element $W_d(t)$ for our model system as a function of time in Fig.~\ref{fig:Framework}(c).
We see that, despite its oscillatory nature, this term decays rapidly for $t \gtrsim 0.2$~ps, corresponding to the period of the fastest mode.
We reiterate that the maximum phonon frequency chosen here is substantially smaller than the typical values seen in ionic conductors, which are closer to 20~THz.
Therefore, the recoil kernel in stiffer real materials is expected to decay even faster.

In the context of ionic motion, the homogeneous portion $\mathbf{r}_H(t) = \sum_n\zeta_j(t)\tensor{m}^{-\frac{1}{2}}\boldsymbol{\varepsilon}_j$ corresponds to thermal vibrations of the lattice and the amplitudes of $\zeta_j(t)=A_j\cos(\Omega_jt+\phi_j)$ need to reflect this fact.
Writing $A_j(n_j)$ explicitly as a function of the quantum excitation level, we have

\begin{align}
	\langle \mathrm{Re}[\zeta_{j}(t)]^2\rangle
 =&
    \oint \frac{d\phi_j}{2\pi}
	\frac{\sum_n A_j^2(n)\cos^2(\Omega_j t + \phi_j) e^{-n\Omega_j / \Omega_T}}{\sum_ne^{-n\Omega_j / \Omega_T}}
	\nonumber
	\\
	=&
   \frac{1}{2}
	\frac{\sum_n A_j^2(n) e^{-n\Omega_j / \Omega_T}}{\sum_ne^{-n\Omega_j / \Omega_T}}
	\,,\label{eqn:zeta_squared}
\end{align}
where $\Omega_T = k_B T / \hbar$ is the thermal frequency.
Using the fact that, for a quantum harmonic oscillator, $\langle \mathrm{Re}[\zeta_{j}(t)]^2\rangle = \frac{\hbar}{\Omega_j}
	\left[n_B(\Omega_j)
	+\frac{1}{2}
	\right]$, where $n_B$ is the Bose-Einstein distribution,
we find that  $A_j(n_j) = \sqrt{n_j + \frac{1}{2}}\sqrt{\frac{2\hbar}{\Omega_j}}$, where $n_j$ is an integer obtained from the probability distribution $e^{-n\Omega_j / \Omega_T}$.
The phase $\phi_j$, on the other hand, is uniformly distributed over $[0, 2\pi]$.

To get a better feel for the framework's thermal motion, we consider a system of $60\times60\times60$ unit cells.
We generate $\mathbf{r}_H(0)$ and $\dot{\mathbf{r}}_H(0)$ for several temperatures by sampling $A_j$'s and $\phi_j$'s and then construct the histograms for the framework mass displacement and speed if Fig.~\ref{fig:Framework}(d) and (e), respectively.

Because it consists of independent harmonics with random phases and thermally-distributed amplitudes, $\mathbf{r}_H(t)$ is a stationary Gaussian process with zero mean and a covariance matrix given by

\begin{align}
   \tensor{C}(t)
 & =
  \langle  \mathbf{r}_H(t)\mathbf{r}_H^\dagger(0) \rangle
  \nonumber
  \\
   & =
  \sum_{l}
 \tensor{m}^{-\frac{1}{2}} \boldsymbol{\varepsilon}_l \boldsymbol{\varepsilon}_l^\dagger\tensor{m}^{-\frac{1}{2}}
  \mathrm{Re}\left[\langle \zeta_l(t)\zeta_l(0)\rangle\right]
  \nonumber
  \\
  & =
  \sum_{l}
  \tensor{m}^{-\frac{1}{2}} \boldsymbol{\varepsilon}_l \boldsymbol{\varepsilon}_l^\dagger\tensor{m}^{-\frac{1}{2}}
   \frac{\hbar\cos(\Omega_l t)}{2\Omega_l}
    \coth\left(\frac{\Omega_l}{2\Omega_T}\right)
\,.    \label{eqn:Correlation}
\end{align}
Diagonal elements of $\tensor{C}(0)$ give the variance of the framework atoms' displacements along individual Cartesian directions.
Multiplying the summand in the penultimate line of Eq.~\eqref{eqn:Correlation} by $\Omega_l^2$ gives the covariance matrix for velocities.
The total displacement probability distribution is given by the Maxwell-Boltzmann form $4\pi r^2 \exp(-r^2 / 2\sigma^2)/(2\pi\sigma)^{3/2}$, where $\sigma$ is the square root of $\tensor{C}(0)$'s diagonal element.
Similarly, we compute the standard deviation for the velocity.
In Fig.~\ref{fig:Framework}(d) and (e), we see excellent agreement between simulated histograms and analytic forms of the distributions over a range of temperatures.

Finally, to describe the interaction between the framework and the particle, we follow our earlier work~\citep{Rodin2022a, Mahalingam2023, Mahalingam2023a, Rodin2024b} and adopt a simplified form by assuming that $U$ is given by the sum of pairwise terms coupling the mobile particle and individual framework masses.
For the pairwise term, we use the simplest physically-motivated form: a screened Coulomb interaction $U(x) = U_0 \exp(-|x|/\lambda) / |x|$ that gives the correct diverging behavior as the separation goes to zero.
In \emph{ab initio} NEB calculations using periodic boundary conditions, the interaction between image charges becomes negligible with supercells larger than a few unit cells in each direction.
This observation indicates that a realistic interaction must be screened on a comparable length scale---unscreened Coulomb interactions are too long-range.

\section{Time-locality validation}
\label{sec:Time_locality_validation}

\begin{figure}
    \centering
    \includegraphics[width=8.5cm]{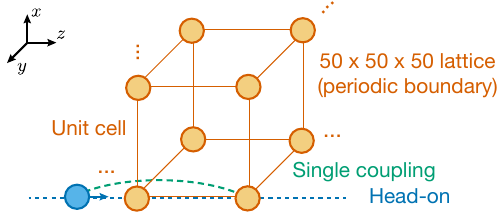}
    \caption{\emph{Schematic of the transport model.} In a cubic lattice of $50\times 50\times 50$ masses with periodic boundary conditions, the mobile particle interacts via a head-on collision with a single mass (shown here as part of a unit cell). We numerically solve for the motion of the mobile particle and the single mass, and use these trajectories to validate the time-local equation.}
    \label{fig:Schematic}
\end{figure}

Once we derived the approximate time-local solution for $\mathbf{r}$ [Eq.~\eqref{eqn:r_by_parts_local}], we wanted to test how reasonable its predictions were.
To this end, we considered a cubic system described in Sec.~\ref{sec:Model_system} composed of $50\times 50\times 50$ masses with periodic boundary conditions and $\mathbf{r}_H = \dot{\mathbf{r}}_H = \mathbf{0}$.
With the framework at rest and undeformed, the mobile particle was launched towards one of the framework masses along one of the edges of a cubic unit cell, set to interact only with its target framework mass, as shown in Fig.~\ref{fig:Schematic}.
We then evolved the system using the full set of equations of motion (full EOM) in Eqs.~\eqref{eqn:Mobile_EOM} and \eqref{eqn:Framework_EOM} using the fifth order Runge-Kutta method, recording the positions and velocities of the particle and the interacting mass.
To assess the validity of the time-local approximation, we compare the ``full EOM'' trajectory $\mathbf{r}$ (the left-hand side of Eq.~\eqref{eqn:r_by_parts_local}) to the ``time-local'' trajectory: the right-hand side of Eq.~\eqref{eqn:r_by_parts_local} calculated by inserting the ``full EOM" $\mathbf{r}$ and $\mathbf{R}$.

\begin{figure*}
    \centering
    \includegraphics[width=\textwidth]{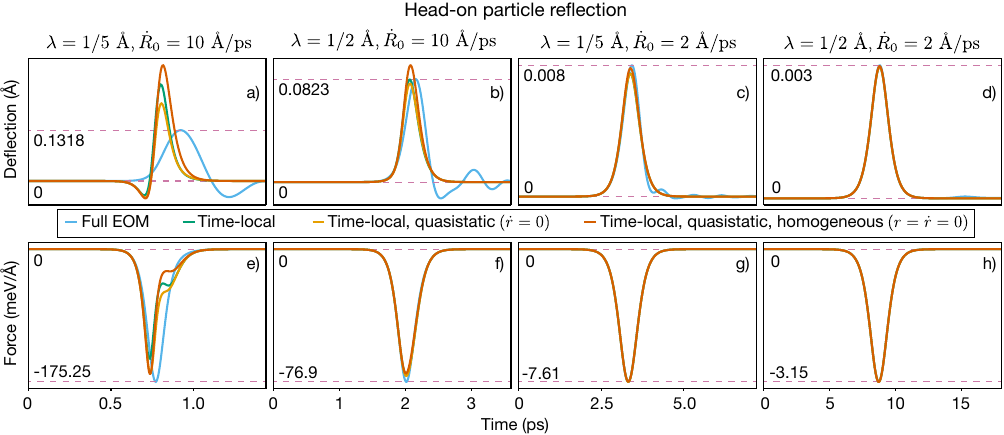}
    \caption{\emph{Validations of the time-local formalism.} We numerically simulated a mobile particle launched towards a framework mass, as shown in Fig.~\ref{fig:Schematic}, with framework properties given in Sec.~\ref{sec:Model_system}. The mobile particle interacts only with its target mass via $U = U_0 e^{-|x|/\lambda}/x$ with $U_0 = 14,000$~meV. Panels (a)--(d) show the deflection of the interacting mass, while (e)--(h) show the force experienced by the mobile particle. The four columns correspond to different combinations of the potential width $\lambda$ and initial particle speed $\dot{R}_0$. Each panel includes four curves corresponding to different levels of approximation. We calculated ``Full" results by solving the equations of motion, Eqs.~\eqref{eqn:Mobile_EOM} and~\eqref{eqn:Framework_EOM}, for the $50\times50\times50$ framework and the mobile particle. We calculated ``Time-local" $r$ by inserting the full-solution $r$, $\dot{r}$, $R$, and $\dot{R}$ into the expression on the right-hand side of Eq.~\eqref{eqn:r_local_single_mass}, and calculated the force using ``time-local'' $r$ and ``full-solution" $R$. For the remaining two solutions, we also used Eq.~\eqref{eqn:r_local_single_mass}, but set $\dot{r}$ or both $r$ and $\dot{r}$ on the right-hand side to zero, in quasistatic and homogeneous approximations, respectively.
    The horizontal dashed lines in all the panels show the amplitude of the maximum displacement and force obtained from the ``Full" solutions.
    }
    \label{fig:Yukawa_Reflection}
\end{figure*}

When we restrict our attention to a single framework mass, Eq.~\eqref{eqn:r_by_parts_local} simplifies dramatically.
In fact, because $\mathcal{L}$ and $\tensor{W}_0(0)$ are diagonal and the particle moves along a high-symmetry direction, we need to consider only a single component of the displacement vector, leading to
\begin{align}
    r &= -w U'(r - R)
    +
    l U''(r - R)\left(\dot{r} -\dot{R}\right)
   \,,
   \label{eqn:r_local_single_mass}
\end{align}
where $w$ and $l$ are the diagonal elements of $\tensor{W}_0(0)$ and $\mathcal{L}$, respectively.

As a benchmark for the time-local approximation, we numerically calculated ``full EOM'' trajectories for several parameter values.
For those parameters where a clear scale exists in common materials, we chose a representative value, while for other parameters, we chose multiple values to illustrate deviations from the ``full EOM'' solutions.
We used an interaction amplitude $U_0 = 14000$~meV, corresponding the mobile and framework ions carrying charges of about $1e$.
For the screening length, we used one of two values: $\lambda = 1/5$~\AA\ or $\lambda = 1/2$~\AA\ so that $U(a)\approx 1.4\times 10^{-3}$~meV or $U(a)\approx 11.6$~meV, respectively.
We also checked the potential energy difference between two mobile particle positions: in the middle of the unit cell and in the middle of one of the faces.
This difference gave a rough estimate of the potential barrier that the particle needs to overcome to move from one unit cell to another.
For $\lambda = 1/5$~\AA, $\Delta U \approx 0.654\text{~meV} - 0.098\text{~meV} = 0.556$~meV and for $\lambda = 1/2$~\AA, $\Delta U \approx 401.4\text{~meV} - 242.1\text{~meV} =159.3$~meV.
The latter energy barrier is similar to typical values found in ionic conductors, so $\lambda=1/2$~\AA\ represents a typical material, while the former barrier is unrealistically low, so $\lambda=1/5$~\AA\ represents a pathological extreme.

We set the starting speed of the mobile particle to either $2$~\AA$/\mathrm{ps}$ or $10$~\AA$/\mathrm{ps}$, giving a total of four simulation runs.
For each run, the particle was initialized $24\lambda$ away from the interacting mass.
The resulting trajectories of the interacting mass obtained from the ``full EOM'' solution are shown in Fig.~\ref{fig:Yukawa_Reflection}(a)--(d) as the light blue curves.

In these trajectories, we see two primary trends.
First, for the same interaction profile, faster-moving particles produce a larger maximum deflection, as expected.
On the other hand, at a given initial speed, the maximum deflection decreases with increasing interaction width.
Physically, for a very narrow potential, the mobile particle delivers an impulse to the framework mass and bounces off before the rest of the framework has time to respond.
Consequently, during the brief time period of contact between the two objects, the framework mass behaves like a free mass.
Conversely, if the potential is very wide, the particle exerts a smaller force on the framework mass over a longer period of time, allowing the neighboring framework masses to respond and provide a restoring force.
This result is consistent with Eq.~\eqref{eqn:r_local_single_mass}, where a wider potential corresponds to a smaller derivative of $U$, decreasing the magnitude of $r$.

After computing the positions and velocities of the particle and the framework mass, we insert them into the right-hand side of Eq.~\eqref{eqn:r_local_single_mass} and plot the resultant ``time-local'' solutions $r$, shown as green curves, along with the ``full EOM'' results.
Naturally, because the time-local formula predicts a deflection only when the mobile particle exerts force on the mass, the curve does not exhibit the oscillations observed in the full solution.
Except for the most pathological case, in Fig.~\ref{fig:Yukawa_Reflection}(a), both the magnitude of the deflection and its time dependence are similar for all the results.

While the deflection of the framework mass is a good check for our solutions, 
the dynamics of the mobile particle are more governed by the force exerted by the framework mass.
Therefore, we plotted the force, $-U'(R - r)$, in Fig.~\ref{fig:Yukawa_Reflection}(e)--(h) calculated from the corresponding deflection in Fig.~\ref{fig:Yukawa_Reflection}(a)--(d).
We see that the difference in the force between the full and time-local results is substantially smaller than the difference in the deflection.
In fact, for all but the most pathological case, the difference is essentially negligible.

Although the calculated displacements of our full and time-local solutions match well, the time-local equation of motion  Eq.~\eqref{eqn:r_by_parts_local} is not solvable on its own.
First of all, even in the single-framework case, Eq.~\eqref{eqn:r_local_single_mass}, there are two unknowns: $r$ and $\dot{r}$.
If we try to circumvent this difficulty by solving for $\dot{r}$ to obtain a differential equation of the form $\dot{r} = f(r)$, this produces a term $r / U''(r - R)$ which diverges when $U''$ goes to zero.
In the more general case of Eq.~\eqref{eqn:r_by_parts_local}, it is impossible to solve for $\nabla_\mathbf{r}dU/dt$ because the matrix $\mathcal{K}\mathcal{L}\mathcal{K}^T$ is singular, so some further simplification is necessary.

To proceed with our under-defined time-local problem, we make use of the fact that the deflection of the framework mass is rather small, so its velocity must also be small: $\dot{r}\rightarrow 0$ on the right-hand side of Eq.~\eqref{eqn:r_local_single_mass}.
Using this quasistatic approximation, we calculated the deflection and force for the same parameters as the other simulations, with the results shown as yellow curves in Fig.~\ref{fig:Yukawa_Reflection}.
We see that this simplification leads to very minor changes in the deflection and even smaller ones in the force.
The main advantage of this simplification is that it turns Eq.~\eqref{eqn:r_local_single_mass} and, by extension, Eq.~\eqref{eqn:r_by_parts_local} into truly time-local transcendental equations for $\mathbf{r}$.
In the presence of thermal motion, this quasistatic approximation amounts to substituting $\dot{\mathbf{r}}\rightarrow \dot{\mathbf{r}}_H$.

Although the quasistatic approximation leads to an important simplification, solving the resulting transcendental equation can still be numerically costly.
Therefore, noticing that the magnitude of the deflection is smaller than the potential width, we set $r \rightarrow 0$ on the right-hand side of Eq.~\eqref{eqn:r_local_single_mass}, which is equivalent to replacing $\mathbf{r}\rightarrow \mathbf{r}_H$ on the right-hand side of Eq.~\eqref{eqn:r_by_parts_local}.
The resulting solutions with a homogeneous-motion approximation are plotted in Fig.~\ref{fig:Yukawa_Reflection} as red curves.
Comparing the various solutions, we see that this dramatic simplification does not introduce a significant difference to the force experienced by the particle.

Based on the results presented in Fig.~\ref{fig:Yukawa_Reflection}, we see that the numerically tractable time-local, quasistatic, homogeneous-motion approximation captures the motion of the framework and the force it exerts on the mobile particle quite well.
Therefore, when exploring dissipation and diffusion in the following sections, we make use of this simplified form of Eq.~\eqref{eqn:r_by_parts_local}.

\section{Dissipation}
\label{sec:Dissipation}

\begin{figure}
    \centering
    \includegraphics[width=\columnwidth]{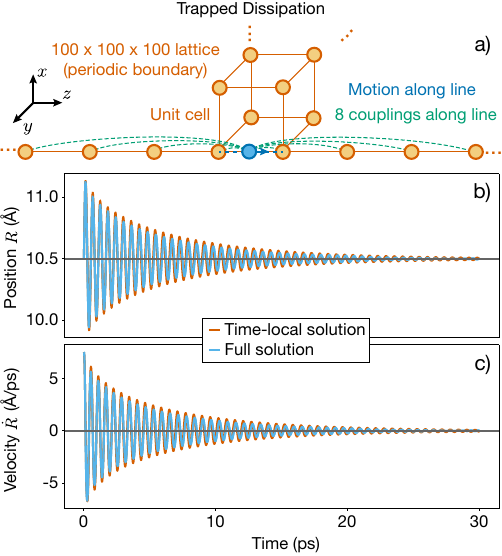}
    \caption{\emph{Dissipation.} (a) A particle constrained to move along one of the edges of the unit cell, interacting with four framework masses ahead of it and four behind via the Yukawa interaction with $\lambda = 1/2$~\AA~and $U_0 = 150$~meV. We computed the full solution using Eqs.~\eqref{eqn:Mobile_EOM} and~\eqref{eqn:Framework_EOM} using the fifth-order Runge-Kutta method with $\delta t = 5\times10^{-3}$~ps. We also computed the time-local result using fifth-order Runge-Kutta from Eqs.~\eqref{eqn:EOM_Simple}--\eqref{eqn:R_eff} with $\mathbf{r}_H = \dot{\mathbf{r}}_H = \mathbf{0}$. The calculated positions and speeds are given in (b) and (c), respectively, showing a very good agreement between the two methods.}
    \label{fig:Trapped_Dissipation}
\end{figure}

Employing the time-local, quasistatic, homogeneous-motion approximation, we set $\dot{r}$ and $r$ to zero on the right-hand side of Eq.~\eqref{eqn:r_local_single_mass} (as in Sec.~\ref{sec:Time_locality_validation}), which yields
\begin{align}
       r
    &= 
    wU'( R)
    -
   l U'' (R) \dot{R}
   \,.
   \label{eqn:r_local_approximate}
\end{align}
Since we already assumed the deflection is small, we employed a series expansion for the force exerted on the mobile particle,
\begin{align}
    M\ddot{R} &= -U'(R - r)
    \nonumber
    \\
    &\approx -U'(R) + U''(R)\left[w
   U'( R)
    -
   l
   U'' (R)
   \dot{R}\right]
   \nonumber
    \\
    &= -U'(R) +  wU''(R)
   U'( R)
    -
   l
   \left[U''(R)\right]^2
   \dot{R} \,.
\end{align}
The first term is essentially the force on the mobile particle from the undeflected framework, the second is a correction due to the framework relaxation, and the third is a force proportional to the speed and directed against the motion---a drag force.

To demonstrate the ability of our formalism to capture drag, we computed two trajectories.
First, we used the full set of equations of motion in Eqs.~\eqref{eqn:Mobile_EOM} and \eqref{eqn:Framework_EOM} for a cubic system of $100\times100\times 100$ masses with periodic boundary conditions and $\mathbf{r}_H = \dot{\mathbf{r}}_H = \mathbf{0}$.
Second, we used the time-local, quasistiatic, homogeneous-motion approximation of Sec.~\ref{sec:Time_locality_validation} given by Eqs.~\eqref{eqn:r_simple}--\eqref{eqn:R_eff} with $\mathbf{r}_H = \dot{\mathbf{r}}_H = \mathbf{0}$.
As before, the calculations were performed using the fifth-order Runge-Kutta method.

To make the results easier to compare, we set up the initial conditions in a manner which produced a one-dimensional path.
In particular, we initialized the mobile particle on one of the edges of a cubic unit cell, halfway between the corners, moving along the edge.
We allowed the particle to interact with eight framework masses: four positioned along the edge ahead of its initial position and four behind [see Fig.~\ref{fig:Trapped_Dissipation}(a)].
We excluded the closer-lying masses not positioned along the edge to speed up the calculation.
Although this restriction does not lead to physically realistic mobile ion trajectories, it does demonstrate dissipation in our model.

Integrating over momenta (Eq.~\eqref{eqn:W_cubic}) is computationally costly, so prior to running the simulation, we precalculated $\tensor{W}_{j-k}(0)$  for all index pairs $j,k$ that are required to connect all the framework masses that interact with the particle at any given time.

For this demonstration, we employed a screened Coulomb interaction with $\lambda = 1/2$~\AA~with $U_0=150$~meV.
We chose this smaller interaction strength to lead to more oscillations before the particle came to rest.
Additionally, because we introduced the particle at a highly energetic position, we wanted to avoid the effects of the boundary term that we dropped from Eq.~\eqref{eqn:r_by_parts}.
In Figure~\ref{fig:Trapped_Dissipation}, we see the position and velocity of the particle computed using the time-local formalism and the full system of equations.
The two solutions show a very similar decay profile, with the full solution decaying slightly faster.
A tiny phase difference is visible at later times, which we attribute to the framework motion present in the full-solution case and absent from the time-local approach.
In short, dissipation is not affected much by our simplifications, even in a more realistic scenario.

\section{Diffusion}
\label{sec:Diffusion}

\begin{figure}
    \centering
    \includegraphics[width=\columnwidth]{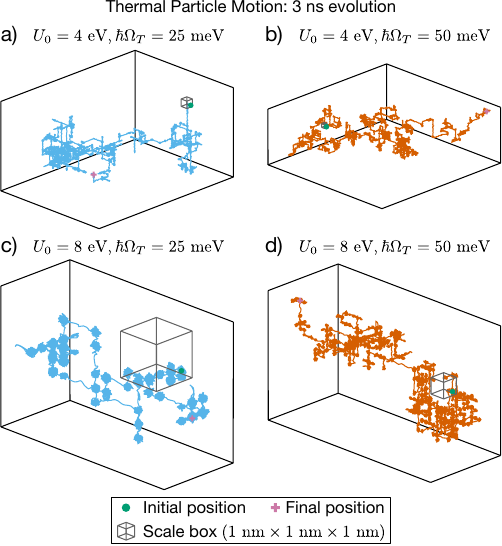}
    \caption{\emph{Time-local diffusion simulations.} Particle trajectories obtained using Eqs.~\eqref{eqn:EOM_Simple}--\eqref{eqn:R_eff} at different temperatures and interaction strengths. For all four cases, $\lambda = 1/2$~\AA. Circles (crosses) denote starting (final) positions. Scale boxes of  $1\text{~nm}\times 1\text{~nm} \times 1$~nm are shown in grey.
    In each case, the underlying lattice geometry is evident in the trajectory, showcasing both motion within unit cells and transport between them.}
    \label{fig:Diffusion}
\end{figure}

Finally, we introduced thermal motion to the framework to check its influence on mobile particle motion.
The computational procedure was essentially identical to the previous section, but with a preliminary procedure and two additional steps.
Prior to the calculation, we generated a set of mode excitation levels $n_j$ and corresponding phases $\phi_j$.
For the first additional step during the trajectory simulation, whenever we needed to calculate the force exerted on the mobile particle, we began by determining which unit cell the mobile particle was in.
Second, assuming that the particle interacts only with the eight framework ions at the corners of this unit cell, we used the pre-generated harmonic amplitudes and phases to calculate $\mathbf{r}_H$ and $\dot{\mathbf{r}}_H$ for those ions.
After that, we proceeded as before by using Eqs.~\eqref{eqn:EOM_Simple}--\eqref{eqn:R_eff} to compute the force experienced by the particle.
This approach led to a thermal trajectory with correct statistical properties while keeping the number of computed framework displacements manageable.
To generate the trajectories, we used $20\times20\times20$ points for each of the three branches the phonon Brillouin zone so that, after discarding the zero-momentum points, we used 23997 modes.

For these calculations, we set $\lambda = 1/2$~\AA, and $U_0 = 4$~eV or 8~eV.
In the unrelaxed configuration, these $U_0$'s resulted in energy barriers between two neighboring unit cells of about $45.5$~meV and $91$~meV, respectively.
If we allowed the lattice to relax, the barriers decreased to $42.3$~meV and $79.3$~meV, respectively.
Although these values are lower than typical energy barriers found in solid ionic electrolytes ($\approx 150$--300~meV for good ionic conductors), they have the correct order of magnitude.
Initially, we chose two temperatures $\hbar\Omega_T = 25$~meV or 50~meV and, with two values of $U_0$, computed four random-walk trajectories spanning 3~ns with $\delta t = 5\times 10^{-3}$~ps, shown in Fig.~\ref{fig:Diffusion}.

\begin{figure}
    \centering
    \includegraphics[width=\columnwidth]{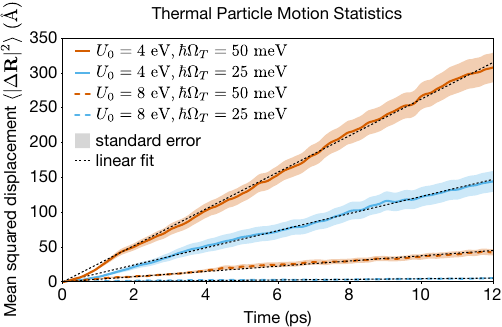}
    \caption{\emph{Mean squared displacement.} For four different combinations of system parameters, we computed particle motion for a total duration of $3$~ns, then divided the trajectory into 12~ps segments. From these segments, we computed the mean and standard error of the squared displacement, depicted with the colored curves and bands, respectively. In all cases, for roughly the first ps, $\langle|\Delta \mathbf{R}|^2\rangle$ grows quadratically with time, due to essentially ballistic particle motion \emph{within} a single unit cell. For longer times, the growth is linear, as shown by agreement with a fit.}
    \label{fig:Variance}
\end{figure}

In line with physical intuition, weaker interactions $U_0$ and higher temperatures $\hbar\Omega_T$ make it easier for the particle to escape the local energy minima, resulting in a longer path.
To quantify the particle's diffusive behavior, we split each trajectory into 250 segments of 12~ps duration each, calculated the squared displacement $|\Delta \mathbf{R}|^2$ as a function of time for each of these segments, then averaged the 250 results to compute the mean $\langle|\Delta \mathbf{R}|^2\rangle$ for each of the four trajectories, shown in Fig.~\ref{fig:Variance}.
We observe that $\langle|\Delta \mathbf{R}|^2\rangle \propto t$, as expected for a diffusive process, where the proportionality constant is $6D$ for 3-dimensional diffusion coefficient $D$.
For the $U_0=8$~eV, $\hbar\Omega_T = 25$~meV case, the simulation time was not long enough to show diffusive motion, as is evident from Fig.~\ref{fig:Diffusion}(c), so the slope of $\langle|\Delta \mathbf{R}|^2\rangle$ does not accurately reflect $D$.

\begin{figure}
    \centering
    \includegraphics[width=\columnwidth]{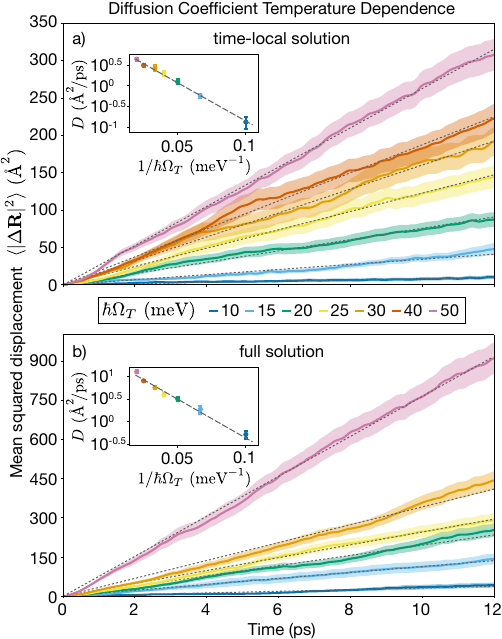}
    \caption{\emph{Temperature dependence of diffusivity.} Main figures: mean square displacement for a mobile particle as a function of time at different temperatures for $U_0 = 4$~eV, $\lambda = 1/2$~\AA. In a), we used the time-local solution, while in b) we used the full solution. In all cases, the initial growth is quadratic, and later evolution is linear, allowing us to extract the slope $6D$ from a fit. Insets: diffusivity $D$ as a function of temperature demonstrates Arrhenius behavior, where the slope of the fit is $-E_a$, and the offset gives $D_0$.}
    \label{fig:Diffusivity}
\end{figure}

To investigate the temperature dependence of diffusivity, we simulated trajectories at several temperatures for $U_0 = 4$~eV.
As above, we partitioned each 3-ns trajectory into 250 segments and used them to calculate the mean square displacement $\langle|\Delta \mathbf{R}|^2\rangle$ as a function of time, shown in Fig.~\ref{fig:Diffusivity}, for both time-local and full solutions.
We excluded the first 1~ps from each segment to reduce the effect of motion \emph{within} a single unit cell.
We fit the resulting $\langle|\Delta \mathbf{R}|^2\rangle$, weighted by its uncertainty, to find its slope for each temperature.
Recalling that, in 3D diffusion, the variance of displacement as a function of time is $6Dt$, we divided the resultant slopes by 6 to yield the diffusion coefficients $D$.
We plot the temperature dependence of the diffusion coefficients in the insets of Fig.~\ref{fig:Diffusivity}.

Because of the large displacement over the 3 ns window, as seen from Fig.~\ref{fig:Diffusion}, it is not computationally feasible to have a sufficiently large system for a full simulation.
Therefore, while the particle's trajectory is continuous, we use periodic boundary conditions to calculate the forces.

Based on an apparent linear relation between $D$ and $1/\hbar\Omega_T$ on the logarithmic scale for both types of solution, we fit the diffusivities using an Arrhenius form $D= D_0e^{-E_a/\hbar\Omega_T}$ and found, with 95\% confidence, $D_0= 9.8^{+2.4}_{-1.9}~(23.0^{+4.4}_{-3.7})$~\AA$^2/\mathrm{ps}$ and $E_a = 42\pm5~(40\pm 5)$~meV for time-local (full) solutions---the latter are close to the estimated barrier height of $42.3$~meV.
This barrier estimate includes both the particle-framework interaction component of $39.4$~meV, as well as the framework deformation energy of $2.9$~meV.

\section{Discussion}
\label{sec:Discussion}

A key approximation we made in order to obtain a time-local expression was setting the maximum phonon momentum to infinity in Eq.~\eqref{eqn:local_kernel}, turning the response kernel into a Dirac delta function.
In reality, of course, the response kernel decays on time-scales similar to the period of the fastest mode, as seen in Fig.~\ref{fig:Framework}(e).
Fortunately, because the decay is quite fast, if we want to bear a greater computational cost in order to more faithfully capture the finite decay time, we can replace the lower limit of the integral in Eq.~\eqref{eqn:r_by_parts} by $\sim t - 2\pi/\Omega_\mathrm{max}$.
As long as the framework deflection is small enough to warrant the $\mathbf{r}\rightarrow\mathbf{r}_H$ switch on the right-hand side of Eq.~\eqref{eqn:r_by_parts}, we can obtain a tractable quasi-time-local description.

For our calculations, we used a cubic lattice and screened Coulomb interactions, for computational convenience.
However, we chose system parameters to try to yield insight into a variety of (more complicated) realistic systems.
In order to demonstrate diffusion in this suboptimal lattice structure, we made the framework less stiff so that its thermal fluctuations could push the mobile particle out of local energy minima more frequently.
Although a softer lattice reduces the applicability of our time-local approach, where a comparison was possible, time-local results showed a good agreement with the full simulations.
We expect that stiffer bonds in real ionic conductors will lead to a greater accuracy of the formalism for two main reasons.
First, the recoil kernel will decay faster, making the Dirac delta function approximation more accurate.
Second, the deflection of the framework atoms will be suppressed, supporting replacing $\mathbf{r}\rightarrow\mathbf{r}_H$ on the right hand side of Eq.~\eqref{eqn:r_by_parts}.

In a good ionic conductor, current carriers easily escape energy minima and spend a long time delocalized before getting trapped again~\cite{Famprikis2019}.
The energy necessary for a charge carrier to leave a minimum originates from the framework's thermal fluctuations.
Unfortunately, because of the fluctuation-dissipation theorem, increased thermal forces experienced by the mobile particles are necessarily accompanied by an increase in drag.
Our analytical results indicate that the drag is inversely proportional to the framework density and the cube of the speed of sound (a measure of the lattice stiffness).
For an infinitely stiff material, the dissipation term vanishes.
Simultaneously, however, the amplitude of the thermal fluctuations vanishes and the particle never acquires sufficient energy to become delocalized.
Conversely, for a very soft material, both the fluctuations and the drag forces are large, suppressing the particle's motion.
Therefore, it is reasonable to assume that, all else being equal, there is some optimal stiffness that balances the fluctuation and dissipation components, leading to the greatest diffusivity.
Similar logic can be applied to the material density.
We expect the improved efficiency of our approach to facilitate explorations of the parameter phase space to build an intuitive understanding of the role that these characteristics play in ionic conductivity.

In our recent work~\citep{Noori2024}, we showed that diffusion coefficients can deviate from Arrhenius scaling even in the simplest case, where the electrolyte is composed of a single atom.
This deviation is also commonly observed experimentally, yet
in Sec.~\ref{sec:Diffusion}, we found that the diffusion coefficient follows an Arrhenius scaling with an activation energy close to the statically-computed value.

In our simulations, the energy barrier was composed primarily of the particle-framework interaction, while the deformation energy played a much smaller role.
Consequently, for the particle to escape a local minimum, it was sufficient that it have enough energy without relying on a favorable framework configuration when it approached the unit cell's face.
For a particle with Boltzmann energy distribution, the probability that it has sufficient energy $E_a$ is approximately proportional to $e^{-E_a/\hbar\Omega_T}$, giving the Arrhenius form for the probability of the particle's escape.
However, if the relaxed configuration involves a large lattice deformation, particle escape also requires a favorable framework arrangement, leading to a deviation from the Arrhenius form.
The reason for this deviation is the fact that the probability of a particle having sufficient energy and the lattice assuming a configuration that allows the particle to move between energy minima is no longer Boltzmann distributed, as discussed in detail in Ref.~\citep{Noori2024}.

\section{Summary}
\label{sec:Summary}

We have presented a scheme for computing the motion of mobile ions in solid electrolytes using a time-local approximation.
By its construction, this formalism is intermediate between fully-static NEB calculations and time-approximation-free AIMD simulations.
Our approach demonstrates the main features associated with ionic transport: dissipation and diffusion.
The simplified computation procedure makes it possible to perform simulations on time and length scales more similar to those anticipated in devices.
In addition to providing a time-local formulation, we also propose a computationally tractable way to capture the short-time-nonlocal effects.

We envision several natural extensions to this work.
First, although we focused on a single mobile ion, it is straightforward to extend our treatment to multiple current carriers.
We do not expect the formalism pertaining to the response kernel to be significantly altered.
Rather, we would only need to include the interaction between the mobile particles to prevent them from occupying the same energy minimum.
We suspect that simulating ionic motion with multiple ions could reveal subtle correlation effects, as well as a departure from the diffusive behavior, leading to anomalous diffusion.

Next, in the limit of high carrier concentration, the interstitial transport picture changes into a vacancy-carrier current.
We suspect modifying our formulation to describe vacancy (rather than particle) motion would not introduce dramatic changes to the structure of the solutions, since the timescale for the response kernel decay should be similar.

Finally, machine learning (ML) techniques can play a substantial role in the simulations.
For the model system considered here, we took a very simple interaction form $U$.
In reality, of course, the potential profile inside a unit cell is very complex and computing it for an arbitrary position of the interstitial atom is computationally costly.
Fortunately, ML has been successfully used to generate potential profiles much faster than using the traditional \emph{ab initio} methods.

For this work, all computations were performed using {\scshape julia}~\citep{Bezanson2017}.
The plots were made with Makie.jl package~\citep{Danisch2021} using the color scheme designed for colorblind readers~\citep{Wong2011}.
The scripts used for computing and plotting can be found at https://github.com/rodin-physics/cubic-lattice-loss-diffusion.

\acknowledgments

A.~R. acknowledges the support by Yale-NUS College (through Start-up Grant).
B.~A.~O. acknowledges support from the M.~J.~Murdock Charitable Trust, and from the the National Science Foundation through Grant No. PHY-2418777.
A.~U. was supported by the Ministry of Education, Singapore, under its funding for the Research Centre of Excellence Institute for Functional Intelligence Materials, National University of Singapore (I-FIM, project No. EDUNC-33-18-279-V12) and by the National Research Foundation, Singapore under its AI Singapore Programme (AISG Award No: AISG3-RP-2022-028).
A.~M. was supported by the Ministry of Education, Singapore, under its funding for the Research Centre of Excellence Institute for Functional Intelligence Materials, National University of Singapore (I-FIM, project No. EDUNC-33-18-279-V12).


%

\end{document}